\documentclass[sigconf]{acmart}

\usepackage{hyperref}
\usepackage{fancyvrb}

\usepackage{amsmath}
\DeclareMathOperator*{\argmax}{arg\,max}

\DefineVerbatimEnvironment{Code}{BVerbatim}{baseline=t}
\newcommand\Tstrut{\rule{0pt}{2.6ex}}         % = `top' strut
\newcommand\Bstrut{\rule[-0.9ex]{0pt}{0pt}}   % = `bottom' strut

\makeatletter
\newcommand*\dotproduct{\mathpalette\dotproduct@{.5}}
\newcommand*\dotproduct@[2]{\mathbin{\vcenter{\hbox{\scalebox{#2}{$\m@th#1\bullet$}}}}}
\makeatother

\usepackage[font=small,labelfont=bf]{caption}
\usepackage{balance}

\def\code#1{\texttt{#1}}

\usepackage{layouts}

\setcopyright{none} % No copyright notice required for submissions
\acmYear{2018}
\acmPrice{}
\acmDOI{}
\acmISBN{}

\acmConference[DYNAMICS'18]{ACM ACSAC conference}{December 2018}{San Juan, Porto Rico, USA\\ This is a pre-print version*}
\begin{document}
\title{Extending Signature-based Intrusion Detection Systems With Bayesian Abductive Reasoning}

\begin{abstract}
Evolving cybersecurity threats are a persistent challenge for system administrators and security experts as new malwares are continually released. Attackers may look for vulnerabilities in commercial products or execute sophisticated reconnaissance campaigns to understand a target's network and gather information on security products like firewalls and intrusion detection / prevention systems (network or host-based). Many new attacks tend to be modifications of existing ones. In such a scenario, rule-based systems fail to detect the attack, even though there are minor differences in conditions / attributes between rules to identify the new and existing attack. To detect these differences the IDS must be able to isolate the subset of conditions that are true and predict the likely conditions (different from the original) that must be observed. In this paper, we propose a \textit{probabilistic abductive reasoning} approach that augments an existing rule-based IDS (snort \cite{roesch1999snort}) to detect these evolved attacks by (a) Predicting rule conditions that are likely to occur (based on existing rules) and (b) able to generate new snort rules when provided with seed rule (i.e. a starting rule) to reduce the burden on experts to constantly update them. We demonstrate the effectiveness of the approach by generating new rules from the snort 2012 rules set and testing it on the MACCDC 2012 dataset \cite{netresec}. 

% Rules based IDS have inherent limitations where they are constantly updated for new attacks that only vary minutely. We use abductive reasoning to generate new rules from existing rule sets to reduce the burden on experts to constantly update them.
\end{abstract}


\author{Ashwinkumar Ganesan}
\affiliation{%
  \institution{Dept. of Computer Science \& Electrical Engineering,\\
  UMBC, \\
  Maryland, USA}
}
\email{gashwin1@umbc.edu}

\author{Pooja Parameshwarappa}
\affiliation{%
  \institution{Dept. Of Information Systems,\\
  UMBC, \\
  Maryland, USA}
}
\email{poojap1@umbc.edu}

\author{Akshay Peshave}
\affiliation{%
  \institution{Dept. of Computer Science \& Electrical Engineering,\\
  UMBC, \\
  Maryland, USA}
}
\email{peshave1@umbc.edu}

\author{Zhiyuan Chen}
\affiliation{%
  \institution{Dept. Of Information Systems,\\ 
  UMBC, \\
  Maryland, USA}
}
\email{zhchen@umbc.edu}

\author{Tim Oates}
\affiliation{%
  \institution{Dept. of Computer Science \& Electrical Engineering,\\
  UMBC, \\
  Maryland, USA}
}
\email{oates@cs.umbc.edu}

\renewcommand{\shortauthors}{Ganesan et al.}

\maketitle

\section{Introduction}
\label{sec:introduction}
The estimated loss to companies and organizations affected by cyber-crimes is increasing \cite{cybercosts}, with targets being attacked through social media platforms such as Twitter and Facebook. Cybersecurity threats are constantly evolving as adversaries design new ways to defeat existing systems. These threats are of two main types: ones that use components of known threats and integrate them to create a ``new" attack and \textit{zero-day}\footnote{\url{https://en.wikipedia.org/wiki/Zero-day_(computing)}} attacks where the attacker discovers a new vulnerability in the product / system that can be exploited before it can be patched up. Although detecting \textit{zero-day} attacks is the ideal expectation, in reality identifying attacks that are slight modifications of existing attacks can be difficult too. Thus, Intrusion Detection Systems (IDS) must be regularly updated with the latest attacks even though attack patterns differ in only small ways. Consider an example of the \textit{Wannacry} ransomware attack\footnote{\url{https://en.wikipedia.org/wiki/WannaCry_ransomware_attack}}. This malware targeted machines that operated on an older version of Microsoft Windows using a known exploit called EternalBlue\footnote{\url{https://en.wikipedia.org/wiki/EternalBlue}}. An analysis of \textit{Wannacry} revealed it to be similar to previous attacks \cite{wannacrysimilar}. The same is true with another well-known ransomware \textit{ExPetr}\footnote{\url{https://securelist.com/schroedingers-petya/78870/}} and a modified version \textit{Bad Rabbit}\footnote{\url{https://securelist.com/bad-rabbit-ransomware/82851/}}. 

This phenomenon is clearly visible when we look at snort rules that contain signature patterns for various cybersecurity threats. Table \ref{int:snort_example_tbl} shows an example set of snort rules that are similar with their corresponding CVE IDs. Snort rule MS06-040 \cite{CVE-2006-3439} tries to alert administrators to a buffer overflow attack on the Microsoft server service while MS08-067 \cite{CVE-2008-4250} checks for an overflow attack triggered by a specific RPC request. Both rules target the same service but have minor variations to accommodate the different methods used to trigger the buffer overflow attack. 

\begin{table*}[t]
  \centering
  \begin{tabular}{cc}
    \hline
    \textbf{Snort Rule} & \textbf{SIG-ID}\\
    \hline
    \begin{Code}
alert tcp $EXTERNAL_NET any -> $HOME_NET [135,139,445,593,1024:]
(msg:"NETBIOS DCERPC NCACN-IP-TCP srvsvc NetrPathCanonicalize overflow attempt"; flow:established,
to_server;dce_iface:4b324fc8-1670-01d3-1278-5a47bf6ee188;byte_jump: 4, -4,multiplier 2,relative,align,
dce; byte_test:4,>,256,0,relative,dce; metadata: policy balanced-ips drop, policy connectivity-ips
drop, policy security-ips drop, service netbios-ssn; classtype:attempted-admin; sid:7209; rev:13;)
    \end{Code}
    & 7209\Tstrut\Bstrut\\\\
    \hline
    \begin{Code}
alert tcp $EXTERNAL_NET any -> $HOME_NET [135,139,445,593,1024:]
(msg:"NETBIOS DCERPC NCACN-IP-TCP srvsvc NetrpPathCanonicalize path canonicalization stack overflow
attempt "; flow:established,to_server; dce_iface:4b324fc8-1670-01d3-1278-5a47bf6ee188; dce_opnum:31,
32;dce_stub_data; pcre:"/^(\x00\x00\x00\x00|.{4}(\x00\x00\x00\x00|.{12}))/sR";byte_jump:4,-4,
multiplier 2,relative,align,dce;pcre:"/\x00\.\x00\.\x00[\x2f\x5c]/R";metadata:policy balanced-ips
drop, policy security-ips drop, service netbios-ssn;classtype:attempted-admin; sid:14782;rev:12;)
   \end{Code}
   & 14782\Tstrut\Bstrut\\\\
   \hline
  \end{tabular}
  \caption{The table shows two snort rules for a Microsoft remote code execution vulnerability (MS06-040\protect\footnote{\url{https://docs.microsoft.com/en-us/security-updates/SecurityBulletins/2006/ms06-040}}, MS08-067\protect\footnote{\url{https://docs.microsoft.com/en-us/security-updates/SecurityBulletins/2008/ms08-067}}) with a critical priority. Their respective CVE-IDs are 2006-3439 and 2008-4250. The differences between them are in four parameters, namely, \textbf{dci\_iface}, \textbf{pcre}, \textbf{dce\_opum} and \textbf{dce\_stub\_data}. The remaining attributes remain the same.}
  \label{int:snort_example_tbl}
\end{table*}

Intrusion Detection Systems (IDS) are of three types \cite{buczak2016survey}:
\begin{enumerate}
    \item\textbf{Signature-based systems} where the attack patterns \cite{roesch1999snort} are defined. These systems cannot detect \text{zero-day} attacks. As new malware is detected, a customized signature must be designed for each attack (or combined with others depending on the system).
    \item\textbf{Machine learning based systems} are of two types. They detect misuse by classifying the traffic as malicious or benign. Anomaly detection systems try to define ``normal" behavior for each process on a host system or the network. These systems cannot detect a specific attack (like EternalBlue) but are able to detect if anomalous (not necessarily malicious) execution happens. The False Alarm Rate (FAR) can be a challenge with anomaly detection mechanisms.
    \item\textbf{Hybrid systems} that combine both machine learning models with signature-based systems.
\end{enumerate}
While signature-based systems require constant rule updates, a major challenge with data-driven methods is their vulnerability to traffic that is skewed between benign and malicious components. The relevant datasets that are openly available (KDD1999 \cite{cup1999dataset}, MACCDC 2012 \cite{netresec}) have a higher degree of malicious traffic as compared to a live stream where a disproportionately large portion of the traffic is benign. As described above, rule-based systems are unable to counter threats that deviate from pre-defined signatures. We solve this problem by building a model that \textit{abduces} likely missing conditions / antecedents from rules.

\textbf{Abductive Reasoning} is a mechanism for generating an inference that explains given observations with maximum likelihood. It is widely used in a number of tasks such as diagnosis of medical conditions based on observed symptoms in a patient \cite{peng2012abductive} to intrusion detection and hypothesizing intrusion objectives \cite{cuppens2002correlation}.

Consider again, the snort rules in table \ref{int:snort_example_tbl}. The \textit{antecedents} of the rule are the conditions that need to be satisfied so that the alert is generated. Thus, conditions such as the protocol (\code{tcp} / \code{udp}), source / target \code{ip-address}, source / target \code{port number}, \code{dce\_iface}, \code{byte\_test} and \code{byte\_jump} are the antecedents in the rule. Similarly, the \textit{consequent} of a rule can be attributes whose values are changed (when the conditions are met). With snort, each consequent is an alert message. The difference between the two rules are the parameters \textbf{dci\_iface}, \textbf{pcre}, \textbf{dce\_opum} and \textbf{dce\_stub\_data}. 

In this paper, we propose a hybrid system that augments a rule-based system (snort) with a probabilistic abductive reasoning model trained from pre-existing snort rules. It performs two tasks, namely, (a) Predict antecedents / conditions in rules that are likely to occur (or are occurring but remain unobserved as they are part of different snort rule whose conditions have not been satisfied), (b) Generate new rules by predicting a unique combination of antecedents. For example, given the antecedents of first rule (SIG-ID: 13162) in table \ref{int:generating_new_rules}, the model generates a new rule (SID: 250001) that has the target port number changed from \code{[139,445]} to \code{[135,139,445,593,1024]} and drops conditions \code{dce\_stub\_data} and \code{byte\_test}.

The pre-existing snort rules are used to learn the correlation between antecedents / attributes in a rule. Once, the correlations are learned, the model abduces antecedents that are likely occur given a seed (initialization) rule. The antecedents from the seed are used to generate new rules so as to expand the coverage of attacks detected by existing rules. 

This paper is divided into the following sections. Section 2 provides an overview of various machine learning, rule based and hybrid methods for intrusion detection. Also, we discuss abductive reasoning methods. Section 3 describes the Bayesian model and pipeline used to generate snort rules. In section 4, we discuss experiments conducted with snort rules dataset and with the MACCDC 2012 dataset. Section 5 describes future directions for our work.%\footnote{The code will be made publicly available on github.}.

\begin{figure*}[ht]
    \centering
    \includegraphics[scale=0.6]{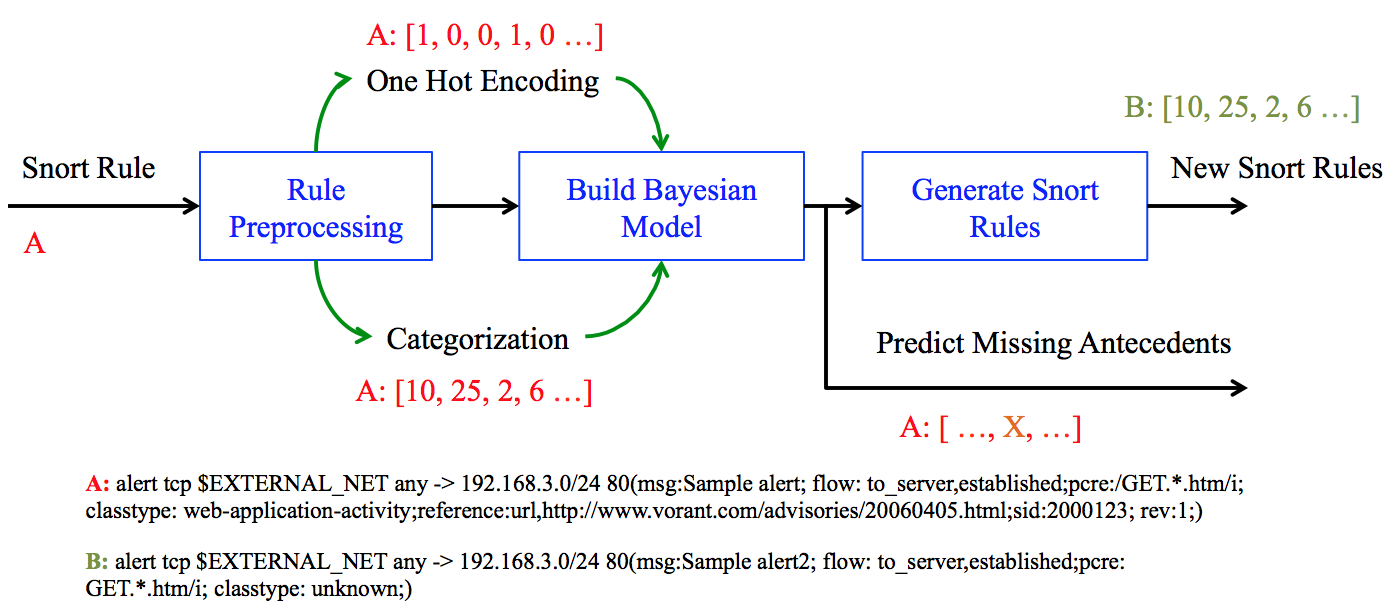}
    \caption{Pipeline to build the Bayesian Abductive Reasoning model. It consists of multiple stages (a) Preprocessing stage where the rule is encoded by categorizing the attributes and then converting them to a one-hot encoded representation (b) Building the Bayesian model (c) Rule generation and attribute prediction. \textbf{A} (in red) is a sample snort alert. \textbf{B} (in green) is the generated snort rule. \textbf{X} represents the missing antecedent in the snort rule \textbf{A}.}
    \label{fig:ab_pipeline}
\end{figure*}

\section{Background \& Related Work}
\label{sec:related_work}
\subsection{Machine Learning for Cybersecurity}
Over the years, a number of techniques have employed machine learning for intrusion detection. Amor et al. \cite{amor2004naive} train a Naive Bayes network to classify attacks and show that it has competitive performance. They compare it against a C4.5 decision tree. Valdes et al. \cite{valdes2000adaptive} construct a Bayesian network (eBayes TCP) for the same purpose. The use of artificial neural networks was explored by Cannady et al. \cite{cannady1998artificial} who trained a neural network to perform multi-category misuse classification. A similar approach was taken by Mukkamala et al. \cite{mukkamala2002intrusion} who compared against a support vector machine. Luo et al. \cite{luo2000mining} learned fuzzy association rules to construct generalized patterns to improve intrusion detection. Genetic algorithms (GA) have been utilized as well \cite{yu2007ensemble}. In inductive learning, rules are induced directly from the training data. One such common inductive learning process is Repeated Incremental Pruning to Produce Error Reduction (RIPPER) \cite{cohen1995fast}. 

Lee et al. \cite{lee1999data} construct a two-stage process where different algorithms (like \textit{frequent episodes}) extract features that RIPPER uses to generate rules. Among more recent methods, Niyaz et al. \cite{javaid2016deep} use a self-taught learning algorithm that combines a sparse autoencoder (unsupervised pre-training) and a multi-layer perceptron (supervised fine tuning) to classify normal and malicious traffic in the KDD99 dataset. Fiore et al. \cite{fiore2013network} use a Discriminative Restricted Boltzmann machine to train a semi-supervised classifier that can detect anomalies in network traffic. The neural network is trained on ``normal" traffic and learns a criteria for normality. Any deviation from normal behavior is flagged as an anomaly. Wang et al. \cite{wang2015applications} similarly try to identify traffic but use a stacked auto-encoder instead while Erfani et al. \cite{erfani2016high} implement a deep belief network (DBN) and fine-tune the model to detect anomalies by using a linear SVM. Ma et al. \cite{ma2016hybrid} combine spectral clustering with a deep neural network to detect attacks. The network is trained in two stages. First, the training data is divided into subset using spectral clustering. The test data is then assigned the pseudo cluster labels depending on the distance of the test datapoint from each cluster. Then, the neural network is trained on the combined set of pseudo labels. Yu et al. \cite{yu2017network} improve on the performance of these network models with a stacked dilated convolutional auto-encoder. Wang et al. \cite{wang2017malware} utilize a convolutional neural network to classify malwares. They subsequently use a convolutional LSTM architecture to learn spatio-temporal features \cite{wang2018hast}. Buczak et al. \cite{buczak2016survey} and Xin et al. \cite{xin2018machine} provide an overview of machine learning and deep learning methods used in cybersecurity.

But training effective machine learning methods is a challenge. Data-driven models require the network traffic on which they are trained to represent the likely distribution of the traffic when they are employed. This forces security analysts to re-train the model on a regular (sometimes daily) basis \cite{buczak2016survey}. Also, there is a lack of good quality labeled data that contains normal and malicious traffic, even though the volume of data available is high. As new attacks are discovered, annotating the data is a continuous and expensive process.  Vu et al. \cite{vu2017deep} try to solve this problem by generating synthetic network traffic using an Auxilliary Generative Adversarial Network. The additional data can be utilized to better train a classifier. In this paper though, we look at methods to enhance existing rule-based systems, benefiting from the rules already created by security analysts.

Although patterns are manually crafted in signature-based IDS, their performance can be improved by automatically generating rules. Gomez et al. \cite{10.1007/978-3-642-21501-8_10} use a pareto-based multi-objective evolutionary algorithm to evolve snort rules. Vollmer et al. \cite{vollmer2011autonomous} try to reduce the effort of creating rules when an intrusion is detected by 
automating the rule creation process. In our research, a Bayesian network is trained on snort rules rather than an evolutionary genetic algorithm.

\subsection{Abductive Reasoning}
As described in the introduction, abductive reasoning is the process of hypothesizing a cause given observed effects. Broadly, abductive reasoning methods can be classified into two types: logic-based \cite{thagard1997abductive} and probabilistic \cite{kate2009rj} methods. Kate et al. \cite{kate2009rj} build a probabilistic abductive reasoning algorithm with a markov logic network. Raghavan et al. \cite{raghavan2010bayesian} designed a Bayesian abductive logic program framework to perform tasks such as plan recognition where the set of observable facts are inadequate to reason deductively. Logic Tensor Networks \cite{serafini2016logic} proposed by Serafini et al. create a single framework to represent first-order predicate logic so as to deductively reason over a knowledge base. 

\section{Problem Definition}
\label{sec:problem_definition}
% \subsection{Scope of Reasoning}
% In this section the definition of \textit{abductive reasoning} is solidified by delineating the scope of the reasoning. In our research, we abduce two specific 

% Extensive work has been conducted on classifying different kinds of cyber attacks. Broadly though, attacks can be be divided into \textit{zero-day} attacks i.e. a new vulnerability is discovered in a system and exploited for a malicious purpose as well as threats with known attack vectors. The latter threats exploit systems whose security has not been updated for a variety of reasons from user indifference to system updates, companies providing incorrect patches for their software / services or slow a system update cycle in an organization even when the relevant patch has been released. Attacks may contain multiple steps too where the attack vector at each step is known but the combination of steps itself might not be known. Our probabilistic abductive reasoning approach is designed to infer these specific threats where the attack vectors are new variations of prior attacks. 

\subsection{Preliminaries}
Consider a set of $n$ rules $R = \{R^1\ ...\ R^n\}$. Let $A = \{A_1\ ...\ A_l\}$ represents the complete set of $l$ antecedents. Each rule $R^i$ has a set of $m$ antecedents given by $R^i = \{a^i_1\ ...\ a^i_m\}$ where $1<m<l$ ($a^i_1$ is a specific value while $A^i_1$ is the category / feature/ variable). The rules can have varying numbers of antecedents. In the following sections, antecedents and attributes are used interchangeably to define the attribute in a snort rule. Snort rules are defined as follows:

\begin{equation}
R^i := \{a^i_1,\ ...,\ a^i_m | a^i_1 \in A^i_1\ ...\ a^i_m \in A^i_m\} 
\end{equation}
is equivalent to:
\begin{equation}
a^i_1 \wedge ... \wedge a^i_m \Rightarrow R^i
\end{equation}

In snort, each rule is considered to be a conjunction of attributes (defined in the alert) / antecedents. Also, snort rules do not use consequent ($R^i$) variables as antecedents (i.e. on L.H.S of the rule). Each component of the rule is associated with an antecedent that is assumed to be a categorical variable with a finite set of possible values. For example, the feature \code{network protocol} has a finite set of values: \code{tcp}, \code{udp} and other protocols. As the number of antecedents in a rule may vary, the variable also has an \code{UNK} token for rules where the antecedent is not present. Each rule represents a particular attack / threat. When $R^i$ is provided as a seed (initialization) rule to the model, it is represented as $O^i$.

% Adding a sample example of a snort here that can be used consistently across the paper.
% 1. Add a single snort rule.
% 2. Add a table of snort antecedents.
%  
\begin{table}[!ht]
  \centering
  \begin{tabular}{p{3cm}|p{4cm}}
    \hline
    \textbf{Antecedent Type} & \textbf{Value}\\
    \hline
    \code{protocol} & \code{tcp}\\
    \code{source IP} & \code{\$EXTERNAL\_NET}\\
    \code{source port} & \code{any}\\
    \code{target IP} & \code{\$HOME\_NET}\\
    \code{target port} & \code{[139,445]}, \code{[135,139,445,593,1024:]}\\
    \code{flow} & \code{established,to\_server}\\
    \code{dce\_iface} & \code{12345678-1234-abcd-ef00 -0123456789ab}\\
    \code{metadata} & \code{policy balanced-ips drop, policy security-ips drop,service netbios-ssn}\\
    \code{dce\_opnum} & \code{0}, \code{1}\\
    \code{byte\_test} & \code{4,>,256,8,relative,dce}, \code{4,>,512,8,relative,dce}\\
    \hline
  \end{tabular}
  \caption{A sample set of antecedents that are extracted from two rules. The rules have all same antecedents except the target port number where one rule checks more ports, \code{byte\_test} and \code{dce\_opnum}.}
  \label{tbl:snort_rules_exp_sample}
\end{table}
To explain the objectives and system architecture, let us consider an example rules set consisting of two rules. Consider the first rule $R^1$ has a SID: 13162 (from table \ref{int:generating_new_rules}). Table \ref{tbl:snort_rules_exp_sample} has the complete set of antecedents identified from the two rules. Thus, the two rules differ in only the \code{target port} numbers, \code{byte\_test} and \code{dce\_opnum}. Given all the antecedents and their possible values, the total number of rules possible are 8 (for each combination \code{target port}, \code{byte\_test} and \code{dce\_opnum}). Let $R^1$ contain the combination \code{target port} = \code{[139,445]}, \code{byte\_test} = \code{4,>,256,8,relative,dce},\code{dce\_opnum} = \code{0} while $R^2$ has \code{target port} = \code{[135,139,445,593,1024:]}, and \code{byte\_test} = \code{4,>,512,8,relative,dce} as well as \code{dce\_opnum} = \code{1}. Let $R^1$ be the seed rule (represented as $O^1$).

\subsection{Definition}
\label{subsec:definition}
In this section, we provide the exact definition of abductive reasoning. Generating a hypothesis rule can be categorized into the following tasks:

\textbf{Abducing Antecedents.} In this task, each hypothesis is considered to be an existing rule with a single antecedent $A^p$ being different. Thus, $A^p \notin \{A^i_j\ |\ A^i_j \in O^i\}$ for given observation $O^i$ (seed rule). 

In case of the example described above, we compute the probability of the hypothesis rule having the \code{target port} = \code{[135,139,445} \code{,593,1024:]} given $R^1$ with \code{byte\_test} = \code{4,>,256,8,relative,} \code{dce} (the combination does not exist in the rules set).

Once the probabilities of all the antecedent values that are not part of the seed rule, are computed, the next step is to select an antecedent as a replacement for its corresponding value in the seed. There are three strategies that can be applied, i.e., choosing the antecedent $A^p$ that has maximum-likelihood, selecting the top $k$ most likely antecedents, or defining a threshold likelihood $t$, above which all antecedents are selected.

\textbf{Abducing Rules.} In this task, each hypothesis is a rule that can have multiple antecedents from the seed rule replaced or inserted. Generating hypotheses where likely antecedents are inserted into the seed rule, has a high computational cost. This is because the possible combinations are exponential. Hence rule abduction is constrained to replacing a set of antecedents in the seed rules only. 

For the previous example, this will lead to a hypothesis rule having \code{target port} = \code{[135,139,445,593,1024:]} and  \ \code{dce\_opnum} = \code{1} if both antecedents are selected.

\begin{table*}[t]
  \centering
  \begin{tabular}{cc}
    \hline
    \textbf{Snort Rule} & \textbf{SIG-ID}\\
    \hline
    \begin{Code}
alert tcp $EXTERNAL_NET any -> $HOME_NET [139,445] (msg:"NETBIOS DCERPC NCACN-IP-TCP spoolss 
EnumPrinters overflow attempt"; flow:established,to_server;
dce_iface:12345678-1234-abcd-ef00-0123456789ab; dce_opnum:0; dce_stub_data; byte_test:4,>,256,8,dce
relative; metadata:policy balanced-ips drop, policy security-ips drop,service netbios-ssn;
reference:bugtraq,21220; reference:cve,2006-5854; reference:cve,2006-6114; reference:cve,2008-0639;
classtype: attempted-admin; sid:13162; rev:9;)
    \end{Code}
    & 13162\Tstrut\Bstrut\\\\
    \hline
    \begin{Code}
alert tcp $EXTERNAL_NET any -> $HOME_NET [135,139,445,593,1024:] (msg: "NETBIOS Generated rule alert
from ID-250001"; metadata:policy balanced-ips drop, policy security-ips drop, service
netbios-ssn; dce_opnum:0;flow:established,to_server;
dce_iface:12345678-1234-abcd-ef00-0123456789ab;byte_test:4,>,256,8,relative,dce; sid:250001;rev:1)
   \end{Code}
   & 250001\Tstrut\Bstrut\\\\
   \hline
  \end{tabular}
  \caption{The table shows a seed rule SID: 13162 and a new generated rule with SID: 250001. The generated rule has all properties of the seed, with one difference: the destination port attribute has been modified from \textbf{[139,445]} to \textbf{[135,139,445,593,1024:]}, thus expanding the rules application to additional ports where an attacker might target the system in the future.}
  \label{int:generating_new_rules}
\end{table*}

\subsection{System Architecture}
% A Bayesian network is implemented for each antecedent in $A$, conditioned on the other antecedents in the rules set $R$.
In this section, each component of the pipeline is described. Figure \ref{fig:ab_pipeline} shows the overall system architecture and how a snort rule is processed. We use Scikit-learn \cite{pedregosa2011scikit} for data preprocessing and constructing the Bayesian network.

\textbf{Rule Preprocessing}.
In this step, the snort rules are parsed and converted to a categorical variable. Snort consists of a fixed set of position attributes, namely, \code{alert}, \code{source IP}, \code{source port number}, \code{destination IP} and \code{destination port number}. The remaining antecedents / attributes are in the form of key-value pairs. We use the keys as nodes in a graphical model and the values represent the set of possible states the variable can take. Thus, a vocabulary of possible values is built for each attribute including attributes such as \code{content} and \code{pcre}. We note that, although, \code{content} and \code{pcre} are strings that can potentially have infinite permutations, in this paper they are considered to have a finite set of possible values bounded by the rules set $R$. Thus each attribute is treated as a categorical variable. Since we construct a multivariate Bayesian model, the attribute values are then converted to one-hot encoded / binary features. In figure \ref{fig:ab_pipeline}, $A$ represents the snort rule (from the rules set $R$), $B$ is the snort rule that is generated and $X$ represents the missing / likely antecedents when a seed rule is provided.

% Add an example case here.

Not all attributes in a snort rule are useful though. A list of excluded attributes is created that contains attributes like \code{sid}, \code{rev} and \code{reference}. The \code{sid} is the signature ID of the rule and \code{rev} is the revision number for the snort rule. \code{Reference} contains external links to information about the attack the snort rule is capturing. Information contained in the \code{reference} attribute can be a URL to CVE description, microsoft security bulletin or another external source. The snort rule for alert \code{A} in figure \ref{fig:ab_pipeline} shows the sample reference. Additionally, attributes that have a constant value like snort rule: \code{alert} are discarded.

\textbf{Building a Bayesian Model}.
Once the snort rules are preprocessed, the one-hot encoded attributes are concatenated to form a feature vector to train the Bayesian network. To train the model and infer efficiently, we assume the attributes to be conditionally independent.
\begin{equation}
    P(a_j\ |\ a_i..a_n) = \argmax_{a_j\in A_j}\prod_{i=1}^{n}P(a_j\ |\ a_i)
\end{equation}
where, $a_j$ is the specific antecedent value ($A_j$ is the categorical variable) to be predicted given the other observed antecedents. While inferring, individual antecedents are not observable. The antecedent is known to be true only when a snort rule generates an alert. Thus, while inferring the values of antecedents, attributes that are not a part of the snort rule are assumed to be \code{UNK}. To generalize the model better for \code{UNK} (unknown) tokens a Laplace smoothing is applied,
\begin{equation}
    P(a_j\ |\ a_i) = \frac{F(a_j,\ a_i) + \alpha}{F(a_i) + \alpha|T|}
\end{equation}
where, $|T|$ represents all the samples in the training set. 

\textbf{Generating Snort Rules}.
% Add an example.
After building the Bayesian network, the model is able to predict the maximum likely antecedent values or the missing attributes when provided with a seed rule $O^i$. As discussed before, we can choose the antecedent either using MLE or the most likely topk values or based on a threshold set manually. Selections based on the MLE can be highly restrictive. Instead, we select all antecedent predictions that are above a threshold $t$. They are combined with the original antecedents of the seed rule in an unordered list. They then form a graph (represented as adjacency matrix) where each value for an antecedent is linked to a value of the next antecedent in the list. We use a depth-first search (DFS) method to generate each possible combinations from the predicted values and eliminate the rules that are copies of the seed. The threshold $t$ has a direct impact on how many rules are generated as it controls values that are predicted.

In the example, let us assume the \code{target port} value \code{[135,139,} \code{445,593,1024:]} and \code{dce\_opnum} value \code{1} have likelihoods greater than the threshold $t$. The additional antecedents are added to the adjacency matrix formed from the attributes in rule $R^1$. We generate all combinations of rules from this graph and eliminate copies of $R^1$. The final rule generated is shown in table \ref{int:generating_new_rules} (rule SIG-ID: 250001).

\subsection{Expanding Features Using Clustering}
\label{subsec:clustering}
Although attributes like \code{content} and \code{pcre} are assumed to be categorical for the purpose of training the model, they can provide insight into rules that are similar based how close the \code{content / pcre} string is to another rule. Thus, clustering can be used to identify similar rules. One of the ways to cluster snort rules is by using hierarchical agglomerative clustering \cite{rokach2005clustering} with a customized Levenshtein distance measure \cite{parameshwarappa2018analyzing} computed in the following manner:
\begin{equation}
    D(r_i,\ r_j) = w_1 \dotproduct KD_{i,j}\ +\ w_2 \dotproduct \sum_{c\in k_i\ \&\ c\in k_j}lev_{c}(r_i^c,\ r_j^c)
\label{eq:levenhstein}
\end{equation}
As seen in equation \ref{eq:levenhstein}, the metric is a weighted distance where, $r_i$ and $r_j$ represent the snort rules, $KD_{i, j}$, the \textit{key} distance, is the symmetric difference between the attributes set in $r_i$ and $r_j$, $lev_{c}(r_i^c,\ r_j^c)$ is the Levenhstein distance between the value of $r_i^c$ and $r_j^c$ (c is a common attribute between $r_i$ and $r_j$). Weights $w_1$ and $w_2$ are hyperparameters.

\section{Experimental Results \& Analysis}
\label{sec:experiment_results}
\subsection{Dataset}
We use two datasets in our experiments. The first is the community edition of snort rules from 2012. The rules are tested on snort version 2.9.21 that was released in 2012. The dataset contains $43792$ rules. A wide variety of rules are available. From these, five categories are selected, namely, \code{special-attacks}, \code{web-misc}, \code{web-cgi}, \code{web-php} and \code{netbios} (selected categories have the large number of rules). Table \ref{tbl:snort_rules_dataset} shows the different types of rules and the number of rules available for each type. As the abduced antecedents or rules generated are specific to each rule set type, the models and experiments are performed independently. The second dataset is the MACCDC 2012 dataset \cite{netresec}. This dataset consists of series of raw pcap files collected from various attack simulations.

\begin{table}[ht]
  \centering
  \begin{tabular}{c|c}
    \hline
    \textbf{Rule Type} & \textbf{Number of Rules}\\
    \hline
    Special Attacks & 902\\
    Web-Misc & 643\\
    Web-CGI & 379\\
    Web-PHP & 201\\
    Netbios & 540\\
    \hline
  \end{tabular}
  \caption{The number of rules for each rule-type in snort that is used in the experiments.}
  \label{tbl:snort_rules_dataset}
\end{table}

\subsection{Abducing Antecedents}
As described in task 1 (sub-section \ref{subsec:definition}), we test whether the model is able to abduce a single missing antecedent. The experiment gives us an idea about the correlations between different attributes that form the rules. The test is conducted in the form of a leave one column out (LOCO) experiment where the column left out is considered missing. Each rule set in table \ref{tbl:snort_rules_dataset} is randomly split into a training and test set with 90\% of data used for training. We perform 10-fold cross-validation to test. Also, the rules are clustered with the customized Levenhstein distance (refer to subsection \ref{subsec:clustering}). After agglomerative clustering is executed, a \textit{clusterID} is assigned to rules that are similar. The clusterID is used as a feature while training the network. The Bayesian model (with and without cluster features) are in green and blue respectively.

\begin{figure*}[ht]
    \centering
    \includegraphics[scale=0.4]{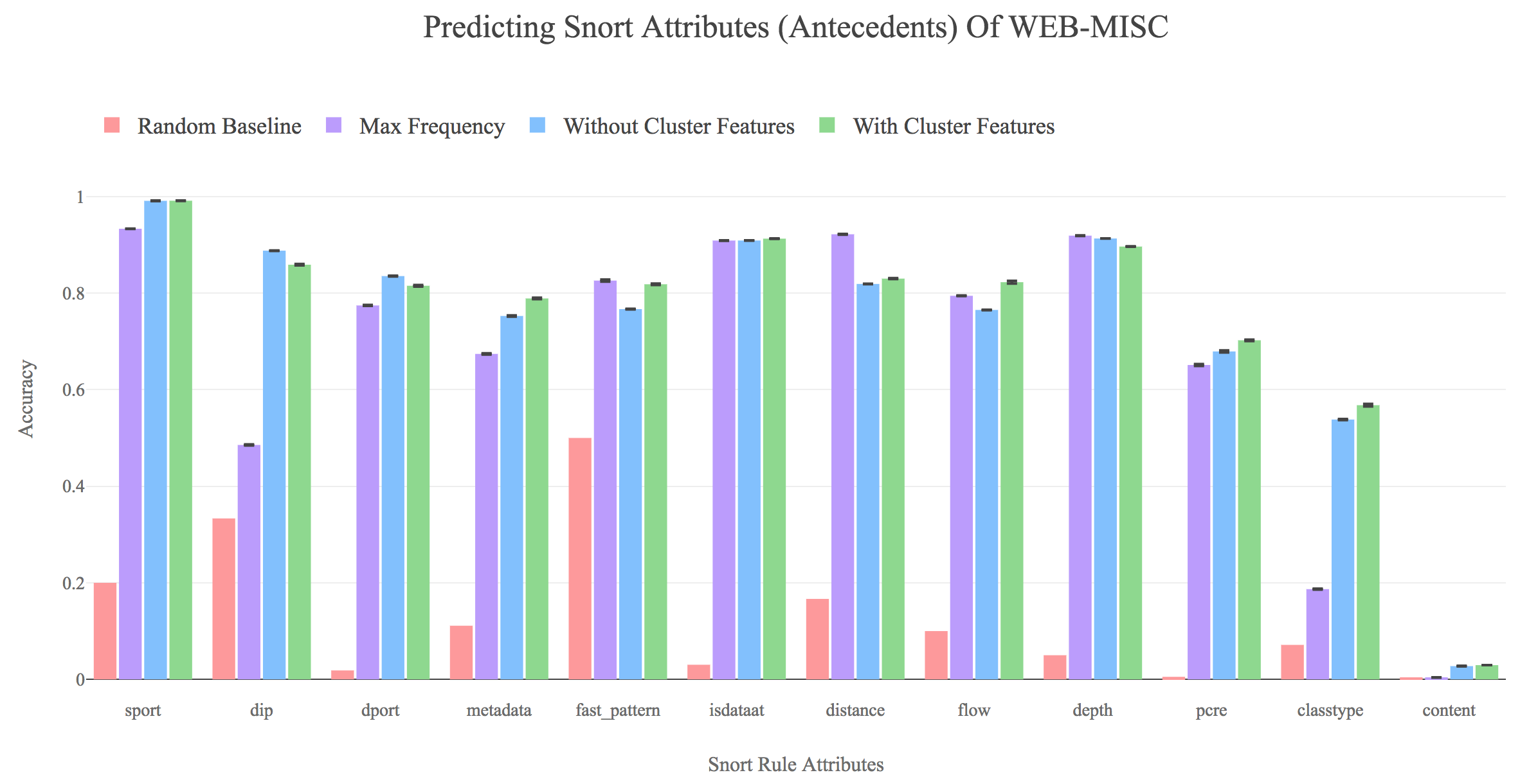}
    \caption{Classification accuracy for each attribute on the \textit{Web-Misc} ruleset.}
    \label{fig:aa_exploit}
\end{figure*}

\begin{figure*}[ht]
    \centering
    \includegraphics[scale=0.4]{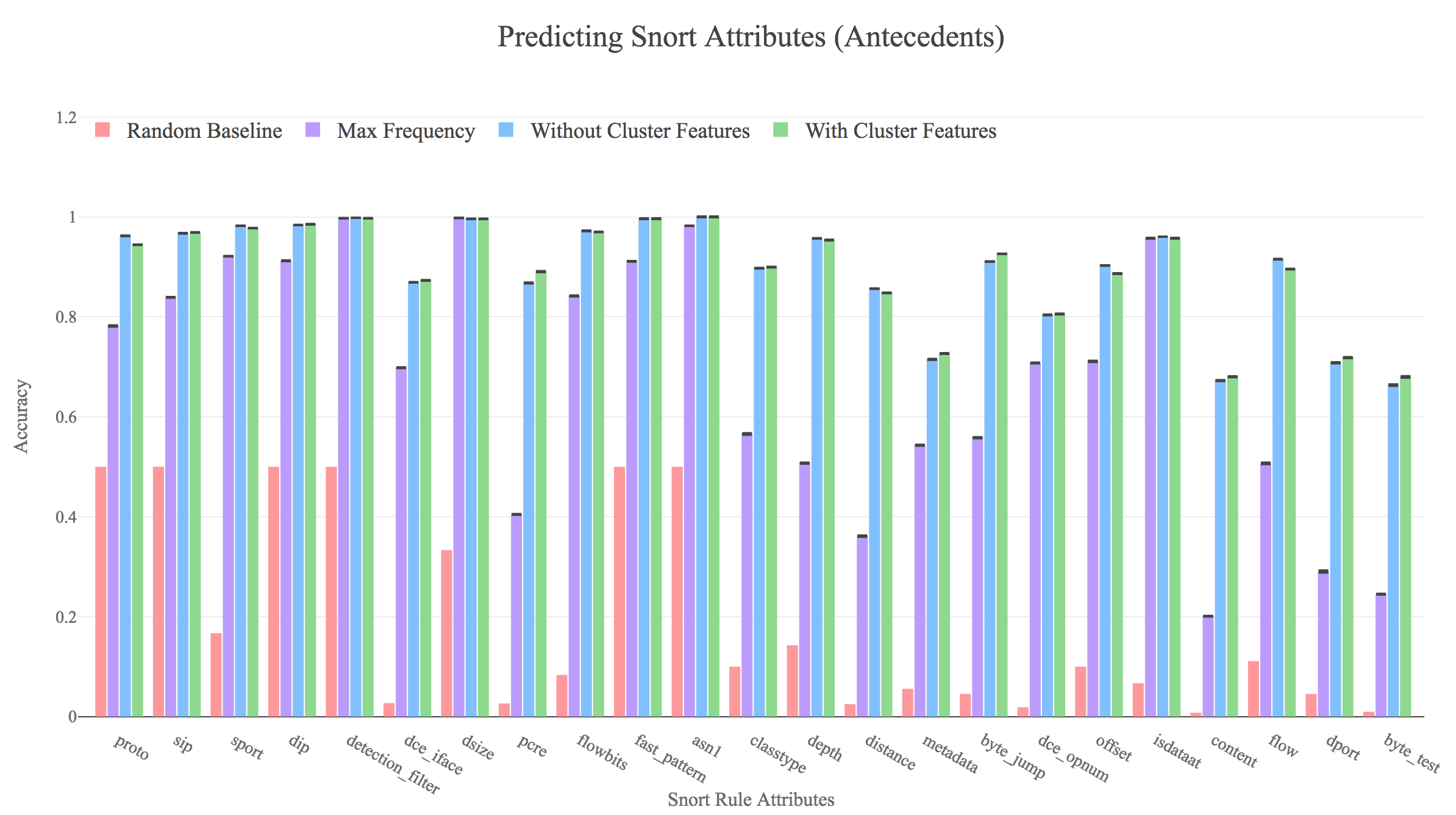}
    \caption{Classification accuracy for each attribute on the \textit{Netbios} ruleset.}
    \label{fig:aa_netbios}
\end{figure*}

\begin{figure*}[ht]
    \centering
    \includegraphics[scale=0.4]{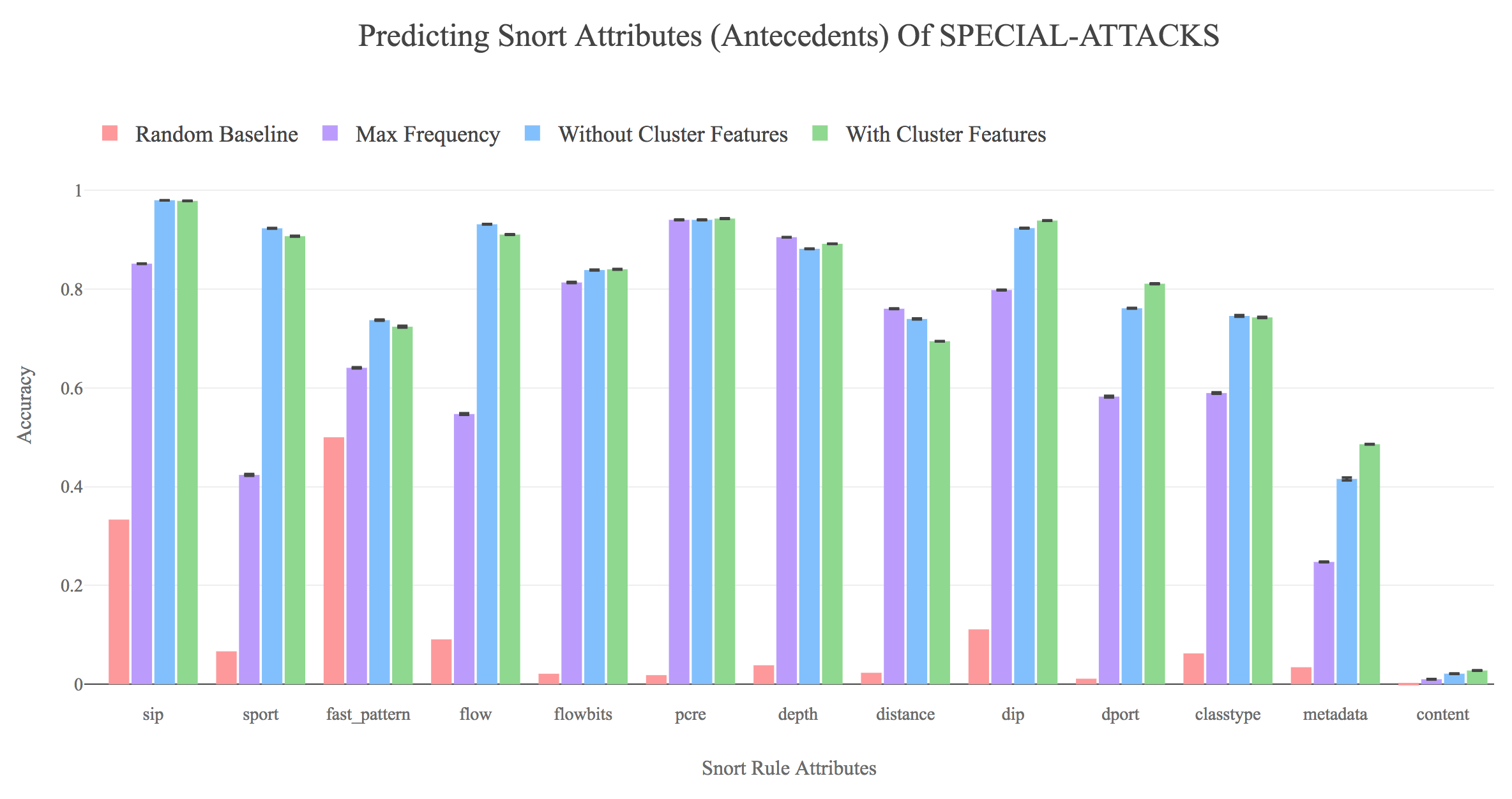}
    \caption{Classification accuracy for each attribute on the \textit{Special Attacks} ruleset.}
    \label{fig:aa_special}
\end{figure*}

\begin{figure*}[ht]
    \centering
    \includegraphics[scale=0.4]{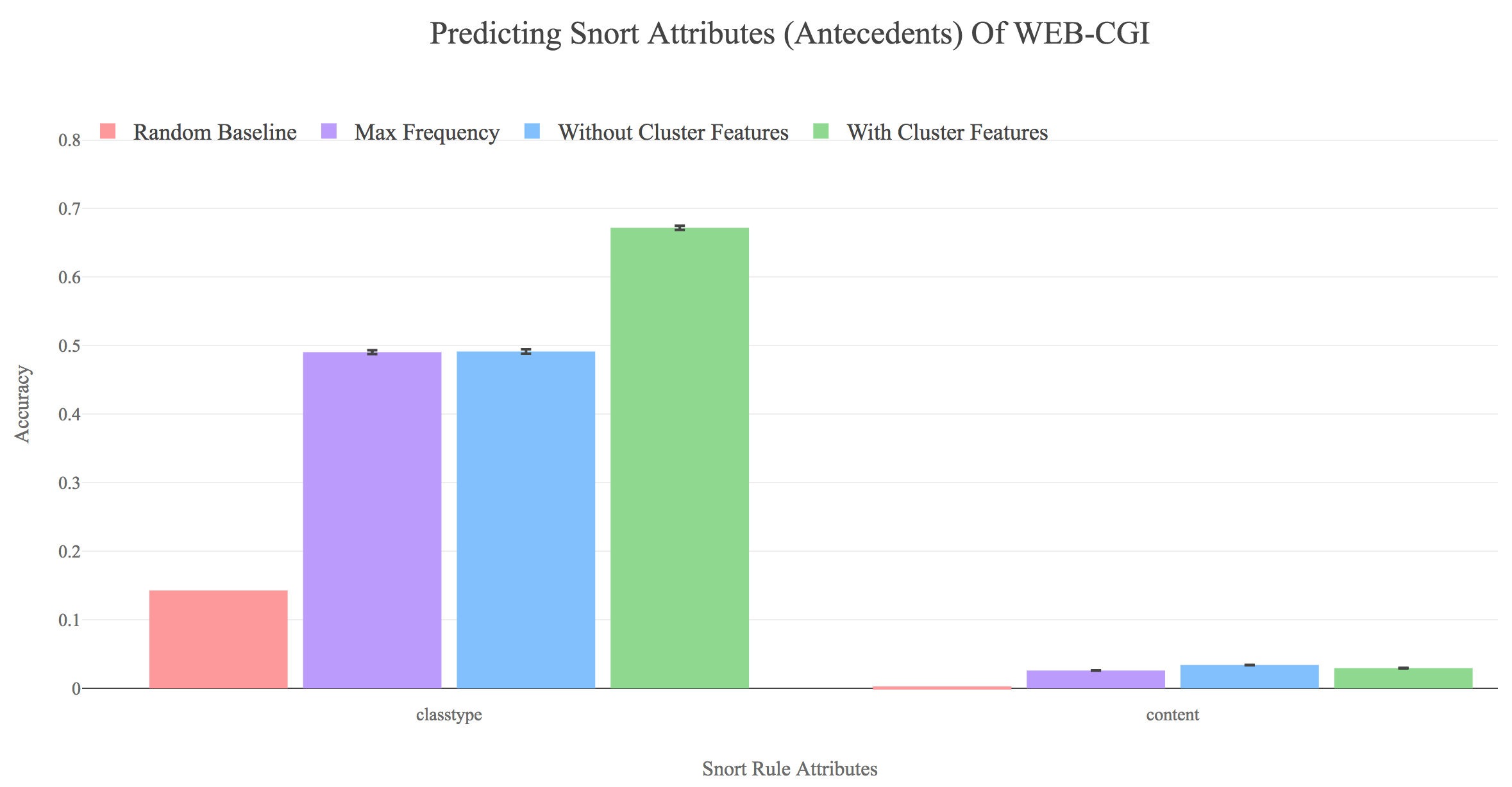}
    \caption{Classification accuracy for each attribute on the \textit{Web-CGI} ruleset.}
    \label{fig:aa_web_cgi}
\end{figure*}

\begin{figure*}[ht]
    \centering
    \includegraphics[scale=0.4]{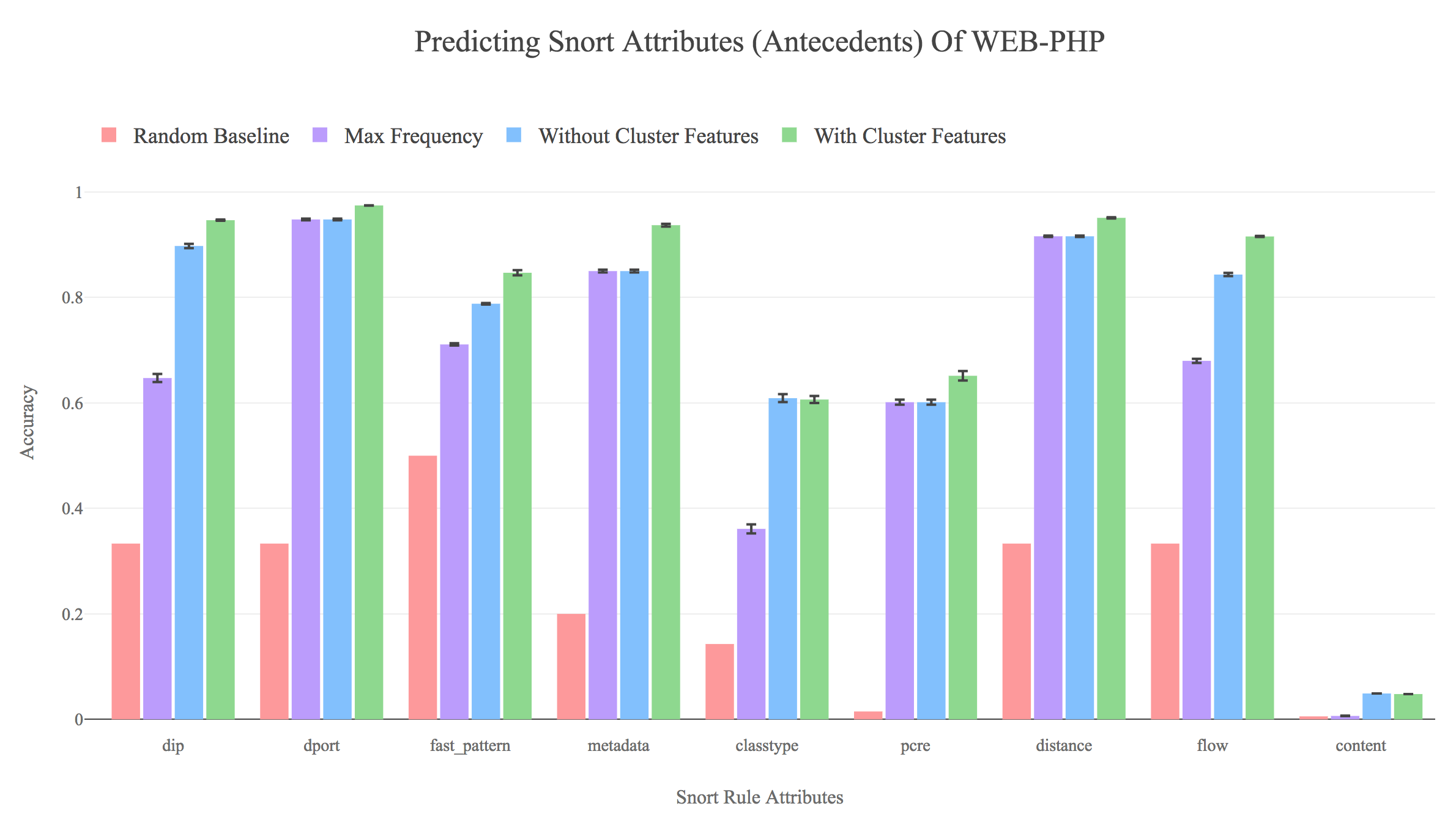}
    \caption{Classification accuracy for each attribute on the \textit{Web-PHP} ruleset.}
    \label{fig:aa_web_php}
\end{figure*}

The performance of the two networks is compared against a random baseline (red) and max frequency classifier (purple) (i.e. a classifier that predicts the same label with the maximum frequency in the training dataset). Figure \ref{fig:aa_exploit}-\ref{fig:aa_web_php} shows the performance of the different classifiers.

\subsection{Analysis \& Discussion}
As seen in Figures \ref{fig:aa_exploit}-\ref{fig:aa_web_php}, the performance of the Bayesian network with and without clusterID feature is better than the max frequency classifier for most attributes. They perform better than the random baseline for all attributes. For attributes such as \code{distance} and \code{depth} in figure \ref{fig:aa_exploit}, \code{detection\_filter} and \code{dsize} in figure \ref{fig:aa_netbios}, \code{depth} in figure \ref{fig:aa_special}, the performance of the max frequency classifier is equal or better than Bayesian models. This is because the classifier performs well when the attribute has a skewed set of values. Consider figure \ref{fig:aa_netbios} that shows the classification performance for Netbios. The Bayesian classifiers have equal accuracy with the maximum frequency classifier for attributes \code{dsize} and \code{detection\_filter}. The following table shows the frequency of individual labels for each of these attributes.
\begin{table}[ht]
    \centering
    \begin{tabular}{c|c}
        \hline
        \textbf{Antecedent} & \textbf{Frequency}\\
        \hline
        \code{UNK} & 538\\
        \code{track by\_dst,count 10,seconds 60} & 2\\
        \hline
    \end{tabular}
    \caption{Frequency of unique \code{detection\_filter} attribute values in the dataset of size 540.}
    \label{tbl:freq1}
\end{table}
\begin{table}[ht]
    \centering
    \begin{tabular}{c|c}
        \hline
        \textbf{Antecedent} & \textbf{Frequency}\\
        \hline
        \code{UNK} & 538\\
        \code{<56} & 1\\
        \code{>100} & 1\\
        \hline
    \end{tabular}
    \caption{Frequency of unique \code{dsize} attribute values in the dataset of size 540. The high frequency of \code{UNK} labels leads a higher performance by the maximum frequency classifier.}
    \label{tbl:freq2}
\end{table}
As seen in the tables \ref{tbl:freq1} and \ref{tbl:freq2}, the majority of the values in \code{detection\_filter} are \code{UNK} tokens (as they are rarely present in snort rules) leading the Bayesian models to perform poorly with respect to other labels. In comparison, the model has a better performance for an attribute like \code{flow} (table \ref{tbl:freq3}) that has a distribution of unique labels that is less skewed.
\begin{table}[ht]
    \centering
    \begin{tabular}{c|c}
    \hline
    \textbf{Antecedent} & \textbf{Frequency}\\
    \hline
    \code{established,to\_server} & 271\\
    \code{UNK} & 119\\
    \code{to\_server, established} & 80\\
    \code{established,to\_client} & 26\\
    \code{stateless} & 24\\
    \code{to\_client,established} & 11\\
    \code{established, to\_server} & 4\\
    \code{to\_server} & 4\\
    \code{established,to\_server,no\_stream} & 1\\

    \hline
  \end{tabular}
  \caption{Frequency of unique \code{flow} attribute values. The unique values have a distribution where the skew is limited. This leads to of a maximum frequency classifier that performs poorly in comparison.}
  \label{tbl:freq3}
\end{table}

\subsection{Qualitative Analysis of New Rules}
To test the quality of the new rules generated (task 2 in sub-section \ref{subsec:definition}), we compare the alerts observed using the seed rule and those generated when the new rules are added to the snort configuration. When snort configuration is updated with them, the seed rule is deactivated. This measures their impact independent of the seed. 

To generate the rules a single seed rule is provided to the model. To generate the rules a threshold posterior probability is defined for each attribute and used to retrieve the \textit{top k} values for each antecedent. We perform our tests with a threshold of $0.01$. 

To compare the alerts generated by both configurations, the pcap file from the MACCDC 2012 dataset is replayed on snort with each individual setting. We analyze if any of the alerts generated while using the new rules are false alarms by associating the timestamps of these alerts for both configurations. Table \ref{tbl:snort_alert_time} shows the timestamps when alerts for seed and new rules are generated. Then, we calculate the number of alerts generated.

% It can be seen that the new rules capture packets at the same time stamp as the original ones, which could mean that the attacks captured by the new rules are valid attacks.

\begin{table*}[ht]
    \centering
    \begin{tabular}{c|c}
    \hline
    \textbf{Time} & \textbf{Snort Alert}\\
    \hline
    \begin{Code}
    3/16-08:48:55.570000  
    \end{Code}
    &
    \begin{Code}
[1:2349:10] NETBIOS DCERPC NCACN-IP-TCP spoolss EnumPrinters attempt [Classification
:Generic Protocol Command Decode] [Priority: 3]
{TCP} 192.168.202.94:52307 -> 192.168.23.100:445    
    \end{Code}
    \Tstrut\Bstrut\\\\
    \hline
\begin{Code}
03/16-08:48:55.570000 
\end{Code}
& 
\begin{Code}
[**] [1:250016:1] NETBIOS Generated rule alert from ID-250016 [**] [Priority: 0]
{TCP} 192.168.202.94:52307 -> 192.168.23.100:445
\end{Code}
\Tstrut\Bstrut\\\\
  \hline
  \end{tabular}
  \caption{The first alert is generated from an existing snort rule (SID:2349) while the second alert is from a snort rule (SID: 2500016) derived using the Bayesian model.}
  \label{tbl:snort_alert_time}
\end{table*}

Since the original rule is a Netbios rule (SID: 13162), we look at the alerts generated for Netbios only. With the original rules, $330$ alerts are generated while with the new rules, $421$ alerts are generated.

%Figure of the logs

\subsection{Impact Of Threshold}
\begin{figure}[ht]
    \centering
    \includegraphics[scale=0.25]{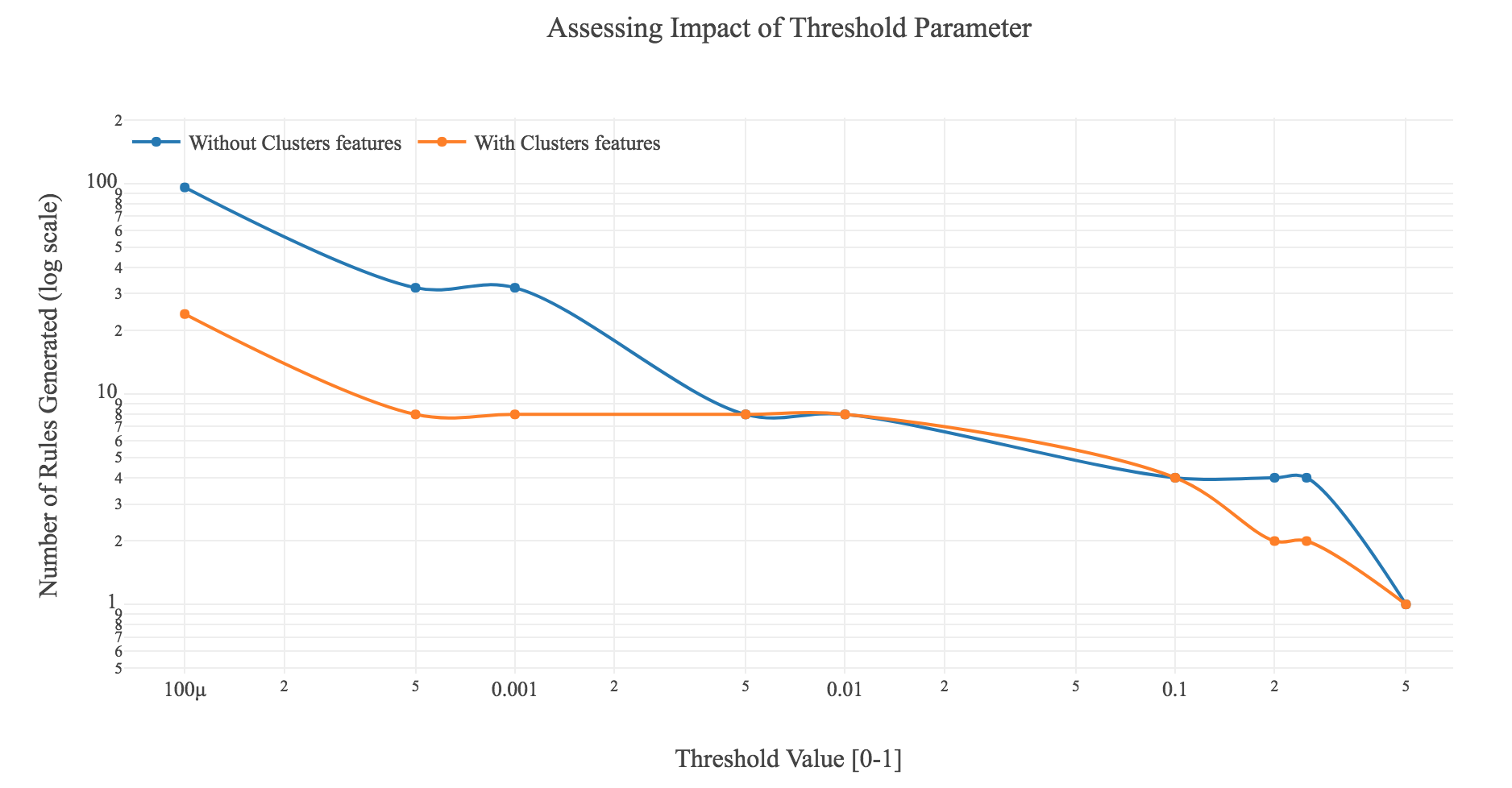}
    \caption{The number of rules generated for different threshold parameters.}
    \label{fig:threshold}
\end{figure}

To test how the threshold affects the rules generated, consider the same seed rule (SID: 13162) as before. We check the rules generated from the seed for varying threshold conditions. The threshold parameter controls the size of the \textit{top k} predicted values for each attribute and thus the types of rules that are generated. A low threshold increases the size of the \textit{top k} list and generates more combinations of rules. On the other hand, with a high threshold, the system may be unable to generate rules at all. To understand the impact of this parameter, we checked the number of rules that are generated for a range of threshold values (as shown in Figure \ref{fig:threshold}).

\section{Conclusion \& Future Work}
In this paper, we show that a Bayesian model trained on snort rules can be utilized to abduce antecedents and to generate a set of new snort rules that are modifications of an existing rule. By treating the missing antecedents either as incomplete or modified conditions, the model provides snort the ability to predict missing antecedents and generate alerts for rules that are likely to be triggered but whose conditions have not been met yet because a potential attacker has modified the threat vector. Also, we show that snort rules are inherently incomplete and designed for specific attacks whose pattern is well established. Abducing new snort rules expands the rules set, prepare the system better for attacks in the future and also provides a measure of the incompleteness of the rule set.

In the future, we will experiment with different graphical and neural network models such as Markov Logic Networks (MLN) \cite{kate2009rj} and Logic Tensor Networks (LTN) \cite{serafini2016logic} as a substitute for our Bayesian approach. Today, LTNs are trained for deductive reasoning but they can be extended to perform abductive reasoning. We can expand the current reasoning approach to abduce rules that have antecedents more than the seed rule and experiment against multiple missing values in the observation as compared to a single missing attribute in this paper. 

Apart from enhancing the current generation of signature-based systems with additional reasoning, the bayesian model provides us with a template for more abstract reasoning. More et. al \cite{more2012knowledge} demonstrate a system that can detect potential attacks by combining information from various ``sensors" on the network i.e. IDS, network traffic analyzers, system logs and so on taking a more holistic view to detect a potential attack. The system can be extended when the ontology is grounded with knowledge about prior attacks. The Unified Cybersecurity Ontology (UCO) does so by combining cybersecurity concepts from multiple known security ontologies like CVE \cite{vulnerabilitiesexposure} and STIX \cite{barnum2012standardizing}. In our view, the bayesian model can be extended to reason with a set of rules designed to operate on such ontologies. Cognitive Cybersecurity System (CCS) \cite{narayanan2018cognitive} is an example of such a system.

The experiments in this paper make an implicit assumption that given the increase in the number of alerts and the timestamp of when the alerts were generated, the rules are detecting potentially new attacks. Our future experiments will analyze the false alarm rates (FAR) for abduced rules.

\section{Acknowledgments}
This research is being conducted in the UMBC Accelerated Cognitive Computing Lab (ACCL) that is supported in part by a gift from IBM Research. We thank the other members of the ACCL Lab for their input, suggestions and guidance in developing this system.

\bibliographystyle{ACM-Reference-Format}
\bibliography{references/security.bib,references/bayes.bib,references/ascps.bib,references/usenix.bib,references/privacy.bib}


\begin{thebibliography}{00}

%%% the Web address.

\ifx \showCODEN    \undefined \def \showCODEN     #1{\unskip}     \fi
\ifx \showDOI      \undefined \def \showDOI       #1{#1}\fi
\ifx \showISBNx    \undefined \def \showISBNx     #1{\unskip}     \fi
\ifx \showISBNxiii \undefined \def \showISBNxiii  #1{\unskip}     \fi
\ifx \showISSN     \undefined \def \showISSN      #1{\unskip}     \fi
\ifx \showLCCN     \undefined \def \showLCCN      #1{\unskip}     \fi
\ifx \shownote     \undefined \def \shownote      #1{#1}          \fi
\ifx \showarticletitle \undefined \def \showarticletitle #1{#1}   \fi
\ifx \showURL      \undefined \def \showURL       {\relax}        \fi
% invisible to TeX
\providecommand\bibfield[2]{#2}
\providecommand\bibinfo[2]{#2}
\providecommand\natexlab[1]{#1}
\providecommand\showeprint[2][]{arXiv:#2}

\bibitem[\protect\citeauthoryear{Amor, Benferhat, and Elouedi}{Amor
  et~al\mbox{.}}{2004}]%
        {amor2004naive}
\bibfield{author}{\bibinfo{person}{Nahla~Ben Amor}, \bibinfo{person}{Salem
  Benferhat}, {and} \bibinfo{person}{Zied Elouedi}.}
  \bibinfo{year}{2004}\natexlab{}.
\newblock \showarticletitle{Naive bayes vs decision trees in intrusion
  detection systems}. In \bibinfo{booktitle}{{\em Proceedings of the 2004 ACM
  symposium on Applied computing}}. ACM, \bibinfo{pages}{420--424}.
\newblock


\bibitem[\protect\citeauthoryear{Barnum}{Barnum}{2012}]%
        {barnum2012standardizing}
\bibfield{author}{\bibinfo{person}{Sean Barnum}.}
  \bibinfo{year}{2012}\natexlab{}.
\newblock \showarticletitle{Standardizing cyber threat intelligence information
  with the Structured Threat Information eXpression (STIX)}.
\newblock \bibinfo{journal}{{\em MITRE Corporation\/}}  \bibinfo{volume}{11}
  (\bibinfo{year}{2012}), \bibinfo{pages}{1--22}.
\newblock


\bibitem[\protect\citeauthoryear{Buczak and Guven}{Buczak and Guven}{2016}]%
        {buczak2016survey}
\bibfield{author}{\bibinfo{person}{Anna~L Buczak} {and} \bibinfo{person}{Erhan
  Guven}.} \bibinfo{year}{2016}\natexlab{}.
\newblock \showarticletitle{A survey of data mining and machine learning
  methods for cyber security intrusion detection}.
\newblock \bibinfo{journal}{{\em IEEE Communications Surveys \& Tutorials\/}}
  \bibinfo{volume}{18}, \bibinfo{number}{2} (\bibinfo{year}{2016}),
  \bibinfo{pages}{1153--1176}.
\newblock


\bibitem[\protect\citeauthoryear{Cannady}{Cannady}{1998}]%
        {cannady1998artificial}
\bibfield{author}{\bibinfo{person}{James Cannady}.}
  \bibinfo{year}{1998}\natexlab{}.
\newblock \showarticletitle{Artificial neural networks for misuse detection}.
  In \bibinfo{booktitle}{{\em National information systems security
  conference}}, Vol.~\bibinfo{volume}{26}. Baltimore.
\newblock


\bibitem[\protect\citeauthoryear{Cohen}{Cohen}{1995}]%
        {cohen1995fast}
\bibfield{author}{\bibinfo{person}{William~W Cohen}.}
  \bibinfo{year}{1995}\natexlab{}.
\newblock \showarticletitle{Fast effective rule induction}.
\newblock In \bibinfo{booktitle}{{\em Machine Learning Proceedings 1995}}.
  \bibinfo{publisher}{Elsevier}, \bibinfo{pages}{115--123}.
\newblock


\bibitem[\protect\citeauthoryear{Competition}{Competition}{2012}]%
        {netresec}
\bibfield{author}{\bibinfo{person}{Mid-Atlantic Collegiate Cyber~Defense
  Competition}.} \bibinfo{year}{2012}\natexlab{}.
\newblock \bibinfo{title}{NETRESEC. 2012. U.S. National CyberWatch Mid-Atlantic
  Collegiate CyberDefense Competition (MACCDC)}.
\newblock \bibinfo{howpublished}{\url{https://www.netresec.com/?page=MACCDC}}.
   (\bibinfo{year}{2012}).
\newblock
\newblock
\shownote{Accessed: 2018-05-09.}


\bibitem[\protect\citeauthoryear{Corporation}{Corporation}{1999}]%
        {vulnerabilitiesexposure}
\bibfield{author}{\bibinfo{person}{MITRE Corporation}.}
  \bibinfo{year}{1999}\natexlab{}.
\newblock \bibinfo{title}{Common Vulnerabilities and Exposures (CVE)}.
\newblock   (\bibinfo{year}{1999}).
\newblock
\showURL{%
\url{http://cve.mitre.org/}}


\bibitem[\protect\citeauthoryear{Cup}{Cup}{1999}]%
        {cup1999dataset}
\bibfield{author}{\bibinfo{person}{KDD Cup}.} \bibinfo{year}{1999}\natexlab{}.
\newblock \showarticletitle{Dataset}.
\newblock \bibinfo{journal}{{\em available at the following website http://kdd.
  ics. uci. edu/databases/kddcup99/kddcup99. html\/}}  \bibinfo{volume}{72}
  (\bibinfo{year}{1999}).
\newblock


\bibitem[\protect\citeauthoryear{Cuppens, Autrel, Miege, and Benferhat}{Cuppens
  et~al\mbox{.}}{2002}]%
        {cuppens2002correlation}
\bibfield{author}{\bibinfo{person}{Fr{\'e}d{\'e}ric Cuppens},
  \bibinfo{person}{Fabien Autrel}, \bibinfo{person}{Alexandre Miege}, {and}
  \bibinfo{person}{Salem Benferhat}.} \bibinfo{year}{2002}\natexlab{}.
\newblock \showarticletitle{Correlation in an intrusion detection process}. In
  \bibinfo{booktitle}{{\em Internet Security Communication Workshop}}.
  \bibinfo{pages}{153--172}.
\newblock


\bibitem[\protect\citeauthoryear{Database}{Database}{2006}]%
        {CVE-2006-3439}
\bibfield{author}{\bibinfo{person}{National~Vulnerability Database}.}
  \bibinfo{year}{2006}\natexlab{}.
\newblock \bibinfo{title}{{CVE}-2006-3439.}
\newblock \bibinfo{howpublished}{Available from MITRE, {CVE-ID}
  {CVE}-2006-3439.}.   (\bibinfo{date}{Aug.} \bibinfo{year}{2006}).
\newblock
\showURL{%
\url{http://cve.mitre.org/cgi-bin/cvename.cgi?name=CVE-2006-3439}}


\bibitem[\protect\citeauthoryear{Database}{Database}{2008}]%
        {CVE-2008-4250}
\bibfield{author}{\bibinfo{person}{National~Vulnerability Database}.}
  \bibinfo{year}{2008}\natexlab{}.
\newblock \bibinfo{title}{{CVE}-2008-4250.}
\newblock \bibinfo{howpublished}{Available from MITRE, {CVE-ID}
  {CVE}-2008-4250.}.   (\bibinfo{date}{Oct.} \bibinfo{year}{2008}).
\newblock
\showURL{%
\url{http://cve.mitre.org/cgi-bin/cvename.cgi?name=CVE-2008-4250}}


\bibitem[\protect\citeauthoryear{Erfani, Rajasegarar, Karunasekera, and
  Leckie}{Erfani et~al\mbox{.}}{2016}]%
        {erfani2016high}
\bibfield{author}{\bibinfo{person}{Sarah~M Erfani}, \bibinfo{person}{Sutharshan
  Rajasegarar}, \bibinfo{person}{Shanika Karunasekera}, {and}
  \bibinfo{person}{Christopher Leckie}.} \bibinfo{year}{2016}\natexlab{}.
\newblock \showarticletitle{High-dimensional and large-scale anomaly detection
  using a linear one-class SVM with deep learning}.
\newblock \bibinfo{journal}{{\em Pattern Recognition\/}}  \bibinfo{volume}{58}
  (\bibinfo{year}{2016}), \bibinfo{pages}{121--134}.
\newblock


\bibitem[\protect\citeauthoryear{Fiore, Palmieri, Castiglione, and
  De~Santis}{Fiore et~al\mbox{.}}{2013}]%
        {fiore2013network}
\bibfield{author}{\bibinfo{person}{Ugo Fiore}, \bibinfo{person}{Francesco
  Palmieri}, \bibinfo{person}{Aniello Castiglione}, {and}
  \bibinfo{person}{Alfredo De~Santis}.} \bibinfo{year}{2013}\natexlab{}.
\newblock \showarticletitle{Network anomaly detection with the restricted
  Boltzmann machine}.
\newblock \bibinfo{journal}{{\em Neurocomputing\/}}  \bibinfo{volume}{122}
  (\bibinfo{year}{2013}), \bibinfo{pages}{13--23}.
\newblock


\bibitem[\protect\citeauthoryear{G{\'o}mez, Gil, Ba{\~{n}}os, M{\'a}rquez,
  Montoya, and Montoya}{G{\'o}mez et~al\mbox{.}}{2011}]%
        {10.1007/978-3-642-21501-8_10}
\bibfield{author}{\bibinfo{person}{J. G{\'o}mez}, \bibinfo{person}{C. Gil},
  \bibinfo{person}{R. Ba{\~{n}}os}, \bibinfo{person}{A.~L. M{\'a}rquez},
  \bibinfo{person}{F.~G. Montoya}, {and} \bibinfo{person}{M.~G. Montoya}.}
  \bibinfo{year}{2011}\natexlab{}.
\newblock \showarticletitle{A Multi-Objective Evolutionary Algorithm for
  Network Intrusion Detection Systems}. In \bibinfo{booktitle}{{\em Advances in
  Computational Intelligence}}, \bibfield{editor}{\bibinfo{person}{Joan
  Cabestany}, \bibinfo{person}{Ignacio Rojas}, {and} \bibinfo{person}{Gonzalo
  Joya}} (Eds.). \bibinfo{publisher}{Springer Berlin Heidelberg},
  \bibinfo{address}{Berlin, Heidelberg}, \bibinfo{pages}{73--80}.
\newblock
\showISBNx{978-3-642-21501-8}


\bibitem[\protect\citeauthoryear{Greenberg}{Greenberg}{2017}]%
        {wannacrysimilar}
\bibfield{author}{\bibinfo{person}{Andy Greenberg}.}
  \bibinfo{year}{2017}\natexlab{}.
\newblock \bibinfo{title}{Wannacry Ransomware Linked suspected North Korean
  hackers}.
\newblock
  \bibinfo{howpublished}{\url{https://www.wired.com/2017/05/wannacry-ransomware-link-suspected-north-korean-hackers/}}.
    (\bibinfo{year}{2017}).
\newblock
\newblock
\shownote{Accessed: 2018-05-09.}


\bibitem[\protect\citeauthoryear{Javaid, Niyaz, Sun, and Alam}{Javaid
  et~al\mbox{.}}{2016}]%
        {javaid2016deep}
\bibfield{author}{\bibinfo{person}{Ahmad Javaid}, \bibinfo{person}{Quamar
  Niyaz}, \bibinfo{person}{Weiqing Sun}, {and} \bibinfo{person}{Mansoor Alam}.}
  \bibinfo{year}{2016}\natexlab{}.
\newblock \showarticletitle{A deep learning approach for network intrusion
  detection system}. In \bibinfo{booktitle}{{\em Proceedings of the 9th EAI
  International Conference on Bio-inspired Information and Communications
  Technologies (formerly BIONETICS)}}. ICST (Institute for Computer Sciences,
  Social-Informatics and Telecommunications Engineering),
  \bibinfo{pages}{21--26}.
\newblock


\bibitem[\protect\citeauthoryear{Kate and Mooney}{Kate and Mooney}{2009}]%
        {kate2009rj}
\bibfield{author}{\bibinfo{person}{Rohit~J Kate} {and}
  \bibinfo{person}{Raymond~J Mooney}.} \bibinfo{year}{2009}\natexlab{}.
\newblock \showarticletitle{RJ: Probabilistic abduction using Markov logic
  networks}. In \bibinfo{booktitle}{{\em In: IJCAI-09 Workshop on Plan,
  Activity, and Intent Recognition}}. Citeseer.
\newblock


\bibitem[\protect\citeauthoryear{Lee, Stolfo, and Mok}{Lee
  et~al\mbox{.}}{1999}]%
        {lee1999data}
\bibfield{author}{\bibinfo{person}{Wenke Lee}, \bibinfo{person}{Salvatore~J
  Stolfo}, {and} \bibinfo{person}{Kui~W Mok}.} \bibinfo{year}{1999}\natexlab{}.
\newblock \showarticletitle{A data mining framework for building intrusion
  detection models}. In \bibinfo{booktitle}{{\em Security and Privacy, 1999.
  Proceedings of the 1999 IEEE Symposium on}}. IEEE, \bibinfo{pages}{120--132}.
\newblock


\bibitem[\protect\citeauthoryear{Luo and Bridges}{Luo and Bridges}{2000}]%
        {luo2000mining}
\bibfield{author}{\bibinfo{person}{Jianxiong Luo} {and}
  \bibinfo{person}{Susan~M Bridges}.} \bibinfo{year}{2000}\natexlab{}.
\newblock \showarticletitle{Mining fuzzy association rules and fuzzy frequency
  episodes for intrusion detection}.
\newblock \bibinfo{journal}{{\em International Journal of Intelligent
  Systems\/}} \bibinfo{volume}{15}, \bibinfo{number}{8} (\bibinfo{year}{2000}),
  \bibinfo{pages}{687--703}.
\newblock


\bibitem[\protect\citeauthoryear{Ma, Wang, Cheng, Yu, and Chen}{Ma
  et~al\mbox{.}}{2016}]%
        {ma2016hybrid}
\bibfield{author}{\bibinfo{person}{Tao Ma}, \bibinfo{person}{Fen Wang},
  \bibinfo{person}{Jianjun Cheng}, \bibinfo{person}{Yang Yu}, {and}
  \bibinfo{person}{Xiaoyun Chen}.} \bibinfo{year}{2016}\natexlab{}.
\newblock \showarticletitle{A hybrid spectral clustering and deep neural
  network ensemble algorithm for intrusion detection in sensor networks}.
\newblock \bibinfo{journal}{{\em Sensors\/}} \bibinfo{volume}{16},
  \bibinfo{number}{10} (\bibinfo{year}{2016}), \bibinfo{pages}{1701}.
\newblock


\bibitem[\protect\citeauthoryear{More, Matthews, Joshi, and Finin}{More
  et~al\mbox{.}}{2012}]%
        {more2012knowledge}
\bibfield{author}{\bibinfo{person}{Sumit More}, \bibinfo{person}{Mary
  Matthews}, \bibinfo{person}{Anupam Joshi}, {and} \bibinfo{person}{Tim
  Finin}.} \bibinfo{year}{2012}\natexlab{}.
\newblock \showarticletitle{A knowledge-based approach to intrusion detection
  modeling}. In \bibinfo{booktitle}{{\em Security and Privacy Workshops (SPW),
  2012 IEEE Symposium on}}. IEEE, \bibinfo{pages}{75--81}.
\newblock


\bibitem[\protect\citeauthoryear{Morgan}{Morgan}{2018}]%
        {cybercosts}
\bibfield{author}{\bibinfo{person}{Steve Morgan}.}
  \bibinfo{year}{2018}\natexlab{}.
\newblock \bibinfo{title}{Cybersecurity facts, figures and statistics}.
\newblock
  \bibinfo{howpublished}{\url{https://www.csoonline.com/article/3153707/security/top-5-cybersecurity-facts-figures-and-statistics.html}}.
    (\bibinfo{year}{2018}).
\newblock
\newblock
\shownote{Accessed: 2018-05-09.}


\bibitem[\protect\citeauthoryear{Mukkamala, Janoski, and Sung}{Mukkamala
  et~al\mbox{.}}{2002}]%
        {mukkamala2002intrusion}
\bibfield{author}{\bibinfo{person}{Srinivas Mukkamala},
  \bibinfo{person}{Guadalupe Janoski}, {and} \bibinfo{person}{Andrew Sung}.}
  \bibinfo{year}{2002}\natexlab{}.
\newblock \showarticletitle{Intrusion detection using neural networks and
  support vector machines}. In \bibinfo{booktitle}{{\em Neural Networks, 2002.
  IJCNN'02. Proceedings of the 2002 International Joint Conference on}},
  Vol.~\bibinfo{volume}{2}. IEEE, \bibinfo{pages}{1702--1707}.
\newblock


\bibitem[\protect\citeauthoryear{Narayanan, Ganesan, Joshi, Oates, Joshi, and
  Finin}{Narayanan et~al\mbox{.}}{2018}]%
        {narayanan2018cognitive}
\bibfield{author}{\bibinfo{person}{Sandeep Narayanan},
  \bibinfo{person}{Ashwinkumar Ganesan}, \bibinfo{person}{Karuna Joshi},
  \bibinfo{person}{Tim Oates}, \bibinfo{person}{Anupam Joshi}, {and}
  \bibinfo{person}{Tim Finin}.} \bibinfo{year}{2018}\natexlab{}.
\newblock \showarticletitle{Cognitive Techniques for Early Detection of
  Cybersecurity Events}.
\newblock \bibinfo{journal}{{\em arXiv preprint arXiv:1808.00116\/}}
  (\bibinfo{year}{2018}).
\newblock


\bibitem[\protect\citeauthoryear{Parameshwarappa, Chen, and
  Gangopadhyay}{Parameshwarappa et~al\mbox{.}}{2018}]%
        {parameshwarappa2018analyzing}
\bibfield{author}{\bibinfo{person}{Pooja Parameshwarappa},
  \bibinfo{person}{Zhiyuan Chen}, {and} \bibinfo{person}{Aryya Gangopadhyay}.}
  \bibinfo{year}{2018}\natexlab{}.
\newblock \showarticletitle{Analyzing attack strategies against rule-based
  intrusion detection systems}. In \bibinfo{booktitle}{{\em Proceedings of the
  Workshop Program of the 19th International Conference on Distributed
  Computing and Networking}}. ACM, \bibinfo{pages}{1}.
\newblock


\bibitem[\protect\citeauthoryear{Pedregosa, Varoquaux, Gramfort, Michel,
  Thirion, Grisel, Blondel, Prettenhofer, Weiss, Dubourg,
  et~al\mbox{.}}{Pedregosa et~al\mbox{.}}{2011}]%
        {pedregosa2011scikit}
\bibfield{author}{\bibinfo{person}{Fabian Pedregosa}, \bibinfo{person}{Ga{\"e}l
  Varoquaux}, \bibinfo{person}{Alexandre Gramfort}, \bibinfo{person}{Vincent
  Michel}, \bibinfo{person}{Bertrand Thirion}, \bibinfo{person}{Olivier
  Grisel}, \bibinfo{person}{Mathieu Blondel}, \bibinfo{person}{Peter
  Prettenhofer}, \bibinfo{person}{Ron Weiss}, \bibinfo{person}{Vincent
  Dubourg}, {et~al\mbox{.}}} \bibinfo{year}{2011}\natexlab{}.
\newblock \showarticletitle{Scikit-learn: Machine learning in Python}.
\newblock \bibinfo{journal}{{\em Journal of machine learning research\/}}
  \bibinfo{volume}{12}, \bibinfo{number}{Oct} (\bibinfo{year}{2011}),
  \bibinfo{pages}{2825--2830}.
\newblock


\bibitem[\protect\citeauthoryear{Peng and Reggia}{Peng and Reggia}{2012}]%
        {peng2012abductive}
\bibfield{author}{\bibinfo{person}{Yun Peng} {and} \bibinfo{person}{James~A
  Reggia}.} \bibinfo{year}{2012}\natexlab{}.
\newblock \bibinfo{booktitle}{{\em Abductive inference models for diagnostic
  problem-solving}}.
\newblock \bibinfo{publisher}{Springer Science \& Business Media}.
\newblock


\bibitem[\protect\citeauthoryear{Raghavan and Mooney}{Raghavan and
  Mooney}{2010}]%
        {raghavan2010bayesian}
\bibfield{author}{\bibinfo{person}{Sindhu Raghavan} {and}
  \bibinfo{person}{Raymond~J Mooney}.} \bibinfo{year}{2010}\natexlab{}.
\newblock \showarticletitle{Bayesian Abductive Logic Programs.}. In
  \bibinfo{booktitle}{{\em Statistical Relational Artificial Intelligence}}.
  \bibinfo{pages}{82--87}.
\newblock


\bibitem[\protect\citeauthoryear{Roesch et~al\mbox{.}}{Roesch
  et~al\mbox{.}}{1999}]%
        {roesch1999snort}
\bibfield{author}{\bibinfo{person}{Martin Roesch} {et~al\mbox{.}}}
  \bibinfo{year}{1999}\natexlab{}.
\newblock \showarticletitle{Snort: Lightweight intrusion detection for
  networks.}. In \bibinfo{booktitle}{{\em Lisa}}, Vol.~\bibinfo{volume}{99}.
  \bibinfo{pages}{229--238}.
\newblock


\bibitem[\protect\citeauthoryear{Rokach and Maimon}{Rokach and Maimon}{2005}]%
        {rokach2005clustering}
\bibfield{author}{\bibinfo{person}{Lior Rokach} {and} \bibinfo{person}{Oded
  Maimon}.} \bibinfo{year}{2005}\natexlab{}.
\newblock \showarticletitle{Clustering methods}.
\newblock In \bibinfo{booktitle}{{\em Data mining and knowledge discovery
  handbook}}. \bibinfo{publisher}{Springer}, \bibinfo{pages}{321--352}.
\newblock


\bibitem[\protect\citeauthoryear{Serafini and Garcez}{Serafini and
  Garcez}{2016}]%
        {serafini2016logic}
\bibfield{author}{\bibinfo{person}{Luciano Serafini} {and}
  \bibinfo{person}{Artur~d'Avila Garcez}.} \bibinfo{year}{2016}\natexlab{}.
\newblock \showarticletitle{Logic tensor networks: Deep learning and logical
  reasoning from data and knowledge}.
\newblock \bibinfo{journal}{{\em arXiv preprint arXiv:1606.04422\/}}
  (\bibinfo{year}{2016}).
\newblock


\bibitem[\protect\citeauthoryear{Thagard and Shelley}{Thagard and
  Shelley}{1997}]%
        {thagard1997abductive}
\bibfield{author}{\bibinfo{person}{Paul Thagard} {and} \bibinfo{person}{Cameron
  Shelley}.} \bibinfo{year}{1997}\natexlab{}.
\newblock \showarticletitle{Abductive reasoning: Logic, visual thinking, and
  coherence}.
\newblock In \bibinfo{booktitle}{{\em Logic and scientific methods}}.
  \bibinfo{publisher}{Springer}, \bibinfo{pages}{413--427}.
\newblock


\bibitem[\protect\citeauthoryear{Valdes and Skinner}{Valdes and
  Skinner}{2000}]%
        {valdes2000adaptive}
\bibfield{author}{\bibinfo{person}{Alfonso Valdes} {and} \bibinfo{person}{Keith
  Skinner}.} \bibinfo{year}{2000}\natexlab{}.
\newblock \showarticletitle{Adaptive, model-based monitoring for cyber attack
  detection}. In \bibinfo{booktitle}{{\em International Workshop on Recent
  Advances in Intrusion Detection}}. Springer, \bibinfo{pages}{80--93}.
\newblock


\bibitem[\protect\citeauthoryear{Vollmer, Alves-Foss, and Manic}{Vollmer
  et~al\mbox{.}}{2011}]%
        {vollmer2011autonomous}
\bibfield{author}{\bibinfo{person}{Todd Vollmer}, \bibinfo{person}{Jim
  Alves-Foss}, {and} \bibinfo{person}{Milos Manic}.}
  \bibinfo{year}{2011}\natexlab{}.
\newblock \showarticletitle{Autonomous rule creation for intrusion detection}.
  In \bibinfo{booktitle}{{\em Computational Intelligence in Cyber Security
  (CICS), 2011 IEEE Symposium on}}. IEEE, \bibinfo{pages}{1--8}.
\newblock


\bibitem[\protect\citeauthoryear{Vu, Bui, and Nguyen}{Vu et~al\mbox{.}}{2017}]%
        {vu2017deep}
\bibfield{author}{\bibinfo{person}{Ly Vu}, \bibinfo{person}{Cong~Thanh Bui},
  {and} \bibinfo{person}{Quang~Uy Nguyen}.} \bibinfo{year}{2017}\natexlab{}.
\newblock \showarticletitle{A Deep Learning Based Method for Handling
  Imbalanced Problem in Network Traffic Classification}. In
  \bibinfo{booktitle}{{\em Proceedings of the Eighth International Symposium on
  Information and Communication Technology}}. ACM, \bibinfo{pages}{333--339}.
\newblock


\bibitem[\protect\citeauthoryear{Wang, Sheng, Wang, Zeng, Ye, Huang, and
  Zhu}{Wang et~al\mbox{.}}{2018}]%
        {wang2018hast}
\bibfield{author}{\bibinfo{person}{Wei Wang}, \bibinfo{person}{Yiqiang Sheng},
  \bibinfo{person}{Jinlin Wang}, \bibinfo{person}{Xuewen Zeng},
  \bibinfo{person}{Xiaozhou Ye}, \bibinfo{person}{Yongzhong Huang}, {and}
  \bibinfo{person}{Ming Zhu}.} \bibinfo{year}{2018}\natexlab{}.
\newblock \showarticletitle{HAST-IDS: learning hierarchical spatial-temporal
  features using deep neural networks to improve intrusion detection}.
\newblock \bibinfo{journal}{{\em IEEE Access\/}}  \bibinfo{volume}{6}
  (\bibinfo{year}{2018}), \bibinfo{pages}{1792--1806}.
\newblock


\bibitem[\protect\citeauthoryear{Wang, Zhu, Zeng, Ye, and Sheng}{Wang
  et~al\mbox{.}}{2017}]%
        {wang2017malware}
\bibfield{author}{\bibinfo{person}{Wei Wang}, \bibinfo{person}{Ming Zhu},
  \bibinfo{person}{Xuewen Zeng}, \bibinfo{person}{Xiaozhou Ye}, {and}
  \bibinfo{person}{Yiqiang Sheng}.} \bibinfo{year}{2017}\natexlab{}.
\newblock \showarticletitle{Malware traffic classification using convolutional
  neural network for representation learning}. In \bibinfo{booktitle}{{\em
  Information Networking (ICOIN), 2017 International Conference on}}. IEEE,
  \bibinfo{pages}{712--717}.
\newblock


\bibitem[\protect\citeauthoryear{Wang}{Wang}{2015}]%
        {wang2015applications}
\bibfield{author}{\bibinfo{person}{Zhanyi Wang}.}
  \bibinfo{year}{2015}\natexlab{}.
\newblock \showarticletitle{The applications of deep learning on traffic
  identification}.
\newblock \bibinfo{journal}{{\em BlackHat USA\/}} (\bibinfo{year}{2015}).
\newblock


\bibitem[\protect\citeauthoryear{Xin, Kong, Liu, Chen, Li, Zhu, Gao, Hou, and
  Wang}{Xin et~al\mbox{.}}{2018}]%
        {xin2018machine}
\bibfield{author}{\bibinfo{person}{Yang Xin}, \bibinfo{person}{Lingshuang
  Kong}, \bibinfo{person}{Zhi Liu}, \bibinfo{person}{Yuling Chen},
  \bibinfo{person}{Yanmiao Li}, \bibinfo{person}{Hongliang Zhu},
  \bibinfo{person}{Mingcheng Gao}, \bibinfo{person}{Haixia Hou}, {and}
  \bibinfo{person}{Chunhua Wang}.} \bibinfo{year}{2018}\natexlab{}.
\newblock \showarticletitle{Machine Learning and Deep Learning Methods for
  Cybersecurity}.
\newblock \bibinfo{journal}{{\em IEEE Access\/}} (\bibinfo{year}{2018}).
\newblock


\bibitem[\protect\citeauthoryear{Yu and Huang}{Yu and Huang}{2007}]%
        {yu2007ensemble}
\bibfield{author}{\bibinfo{person}{Yan Yu} {and} \bibinfo{person}{Hao Huang}.}
  \bibinfo{year}{2007}\natexlab{}.
\newblock \showarticletitle{Ensemble approach to intrusion detection based on
  improved multi-objective genetic algorithm.}
\newblock \bibinfo{journal}{{\em Ruan Jian Xue Bao(Journal of Software)\/}}
  \bibinfo{volume}{18}, \bibinfo{number}{6} (\bibinfo{year}{2007}),
  \bibinfo{pages}{1369--1378}.
\newblock


\bibitem[\protect\citeauthoryear{Yu, Long, and Cai}{Yu et~al\mbox{.}}{2017}]%
        {yu2017network}
\bibfield{author}{\bibinfo{person}{Yang Yu}, \bibinfo{person}{Jun Long}, {and}
  \bibinfo{person}{Zhiping Cai}.} \bibinfo{year}{2017}\natexlab{}.
\newblock \showarticletitle{Network intrusion detection through stacking
  dilated convolutional autoencoders}.
\newblock \bibinfo{journal}{{\em Security and Communication Networks\/}}
  \bibinfo{volume}{2017} (\bibinfo{year}{2017}).
\newblock


\end{thebibliography}
\end{document}